\documentclass[letterpaper]{article} 
\usepackage{aaai2026}  
\usepackage{times}  
\usepackage{pifont}
\usepackage{booktabs}
\usepackage{adjustbox}
\usepackage{nicematrix}
\usepackage[table]{xcolor}
\usepackage{multirow} 
\usepackage{helvet}  
\usepackage{courier}  
\usepackage[hyphens]{url}  
\usepackage{graphicx} 
\usepackage{natbib}  
\usepackage{caption} 
\usepackage{algorithm}
\usepackage{algorithmic}

\usepackage{newfloat}
\usepackage{listings}
\DeclareCaptionStyle{ruled}{labelfont=normalfont,labelsep=colon,strut=off} 
\floatstyle{ruled}
\newfloat{listing}{tb}{lst}{}
\floatname{listing}{Listing}
\title{Tool-RoCo: An Agent-as-Tool Self-organization Large Language Model Benchmark in Multi-robot Cooperation}
\author{
    Ke Zhang\textsuperscript{\rm 1}\thanks{Preprint. Work conducted during the AI research internship at Advanced Technology Center, SONY~(China)}, Xiaoning Zhao\textsuperscript{\rm 1},
    Ce Zheng\textsuperscript{\rm 2},
    Jiahong Ning\textsuperscript{\rm 2},
    Dandan Zhu\textsuperscript{\rm 3}, \\
    Wenqi Zhang\textsuperscript{\rm 4},
    Chen Sun\textsuperscript{\rm 4},
    Toshiharu Sugawara\textsuperscript{\rm 1}
}

\affiliations{
    \textsuperscript{\rm 1}Dept. of Computer Science and Communications Engineering, Waseda University, Tokyo, Japan\\
    \textsuperscript{\rm 2}Dept. of Network Intelligence, Pengcheng Laboratory, Shenzhen, China\\
     \textsuperscript{\rm 3}Dept. of Artificial Intelligence, China University of Petroleum, Beijing, China\\
    \textsuperscript{\rm 4}Advanced Technology Center, SONY (China) Co., Ltd\\
   zhangke@moegi.waseda.jp
}

\usepackage{bibentry}
\makeatletter
\gdef\copyright@on{}
\makeatother

\begin{document}

\maketitle

\begin{abstract}
This study proposes Tool-RoCo, a novel benchmark for evaluating {\em large language models} ~(LLMs) in long-term multi-agent cooperation based on RoCo, a multi-robot cooperative benchmark. Recent research on LLM-based multi-agent systems has relied on predefined orchestration, while ignoring agent autonomy. Tool-RoCo treats other agents as tools and introduces cooperative tools, leveraging tool usage to evaluate multi-agent cooperation and self-organization. Tool usage means that each agent (LLM) selects a tool from a candidate set based on the current state, receives feedback, and adjusts its selection in subsequent rounds. To evaluate different autonomy levels, we propose four LLM paradigms: (1) {\em centralized cooperation}, where a single LLM allocates tools to all agents; (2) {\em centralized self-organization}, where a central LLM autonomously activates agents while keeping others inactive; (3) {\em decentralized cooperation}, where each agent has its own LLM and calls tools based on local information; and (4) {\em self-organization}, where a randomly chosen initial agent can request collaboration, activating additional agents via tool calls. Tool-RoCo includes three multi-robot tasks, SORT, PACK, and CABINET, to measure format and parameter accuracy and agent coordination through tool usage. The results using several LLMs showed that cooperative tools accounted for only 7.09\% of all tools, indicating that LLM-based agents rarely invoked others as assistants. Moreover, activation tools accounted for 96.42\%, suggesting that current LLMs tend to maintain active agents while seldom deactivating them for adaptive coordination. Tool-RoCo provides a systematic benchmark to evaluate LLM autonomy and cooperation in multi-agent tasks. 

\end{abstract}

\begin{links}
    \link{Code}{https://github.com/ColaZhang22/Tool-Roco}
\end{links}

\section{Introduction}

Recent advances in LLM-based multi-agent systems~(MAS) significantly enhance performance across various industries, such as multi-agent pathfinding~\cite{chen2025solving}, multi-turn negotiation~\cite{dong2024multi}, and multi-robotics cooperation (RoCo)~\cite{mandi2024roco,wang2025capabilitiesgpt5multimodalmedical}. In these tasks, LLMs have demonstrated capabilities in high-level planning and task decomposition, making them promising for intricate multi-agent cooperation. However, existing LLM-based MAS benchmarks~\cite{wu2023autogenenablingnextgenllm,chen2024agentverse} mainly rely on predefined characters or fixed cooperative pipelines, leveraging fixed dialogue sequences and characterized prompting engineering to enhance the performance of LLM. It remains unclear how LLMs can coordinate and self-organize with other agents in long-term tasks, limiting the evaluation of their autonomy.

\par
Meanwhile, the tool usage of LLM offers a natural entry point for evaluating LLM cooperation and self-organization abilities. Tool usage enables LLM-based agents to shift from directly solving complex tasks to leveraging external tools, such as functions and APIs, to provide a structured response and feedback for evaluating multi-agent LLM abilities. According to the structured response, current tool usage benchmarks enable the evaluation of LLM by tool selection correctness and parameter filling success in tasks, but do not consider the dynamics of agent-to-agent coordination over long-term interaction environments.

\begin{figure}[h]
    \centering
    \includegraphics[width=1\linewidth]{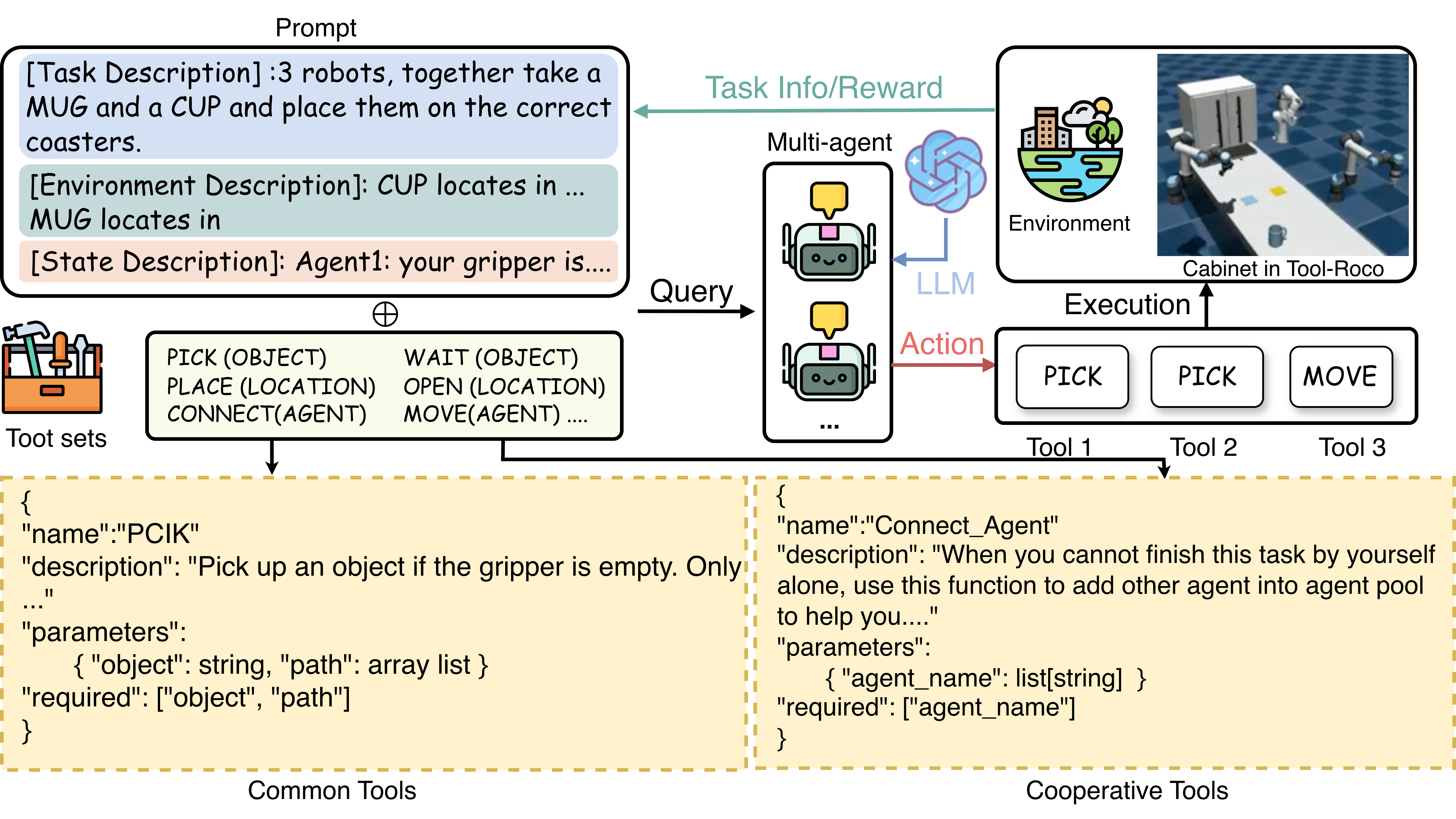}
    \caption{Overview of Tool-RoCo pipeline and two kinds of tools: LLM agents perceive the current state and select appropriate tools to take action or call assistance from other agents. Then, the environment feedback and add the execution result to the next state. }
    \label{fig:overview}
\end{figure}

\par

To address the limitations of current LLM-based multi-agent systems, we propose Tool-RoCo, a long-term multi-robot cooperation benchmark for evaluating the autonomy and coordination capacity of LLM agents. Tool-RoCo adapts three cooperative tasks, CABINET, PACK, and SORT, in RoCo~\cite{mandi2024roco}. Each task requires multi-agent leverage of appropriate tools to cooperatively complete the target, such as picking up a cup and a mug to place them in the correct position. In contrast to prior multi-agent systems that rely on predefined pipelines and fixed characters, Tool-RoCo introduces a novel viewpoint by treating agents as tools. As shown in Fig.\ref{fig:overview}, each LLM agent in Tool-RoCo selects and executes a tool from a candidate tool set based on the current state and then receives feedback from the environment. This loop enables agents to iteratively refine their tool choices over multiple interaction rounds.

\par

Based on this viewpoint, Tool-RoCo defines four types of cooperative paradigms to evaluate various levels of autonomy, as shown in Fig.\ref{fig:method}. First, the lowest autonomous level, the centralized execution paradigm, involves a single LLM that perceives the entire environment state and uniformly selects tools for all agents. The next level, the centralized self-organized paradigm, allows LLM to treat all agents as tools and selectively activate or deactivate agents in each turn. The third level is the decentralized cooperation paradigm. Each agent, equipped with an LLM, is independent of the other agents and possesses an individual tool set to select the appropriate tools. Finally, the self-organization paradigm is based on independent agents in the decentralized cooperation paradigm: a randomly chosen agent is initialized as active, and this agent can dynamically activate or deactivate other agents to self-organize and collaboratively achieve the task goals.

\par
These four paradigms establish a progressive evaluation for multi-agent systems, ranging from tool selection correctness and parameter filling to cooperation and self-organization. This hierarchical paradigm provides a systematic assessment of LLM-based agent autonomy and coordination, offering a comprehensive benchmark for long-term cooperative tasks.
\par
In conclusion, our contribution can be summarized as follows:
\begin{itemize}
\item We introduce Tool-RoCo, an LLM-based multi-agent benchmark for long-term multi-robot cooperation tasks, which contains four cooperation paradigms. Unlike existing tool usage benchmarks, Tool-RoCo first conceptualizes agents as tools and agent uses other agents as tools to measure how LLM agents autonomously cooperate with other agents.
\item Tool-RoCo proposes two novel metrics: self-organization~(SO) and cooperation-tool ratio(CT). Self-organization quantifies the degree to which an agent autonomously seeks assistance from other agents, and the cooperation-tool ratio is the proportion of cooperative tool usage. These two metrics provide a novel evaluation of the autonomy and cooperative capacities of LLM-based agents.
\item Four LLMs are used to test and verify three multi-robot cooperation tasks: CABINET, PACK, and SORT in Tool-RoCo. The results reveal that current LLM agents tend to maintain cooperation (Average SO is 96.42\%) rather than disconnect redundant and unnecessary agents. Notably, the cooperation tool usage ratio(CT) remained relatively low (7.09\%), indicating that the LLM agents preferred to individually attempt various tools rather than call for assistance from other agents.
\end{itemize}

\section{Related Work}

\subsection{Multi-Agent Systems with LLMs}

Current LLM-based multi-agent systems can be divided into two main categories, as shown in Table \ref{tab:benchmarks}. The first~\cite{wu2023autogenenablingnextgenllm} leverages a multi-agent cooperative paradigm to enhance the performance of LLM in conventional tasks, such as mathematics\cite{lei2024macm} and code
~\cite{zhao2025mage}. For instance, MultiAgentBench~\cite{zhu_multiagentbench_2025}, LLM-Coordination~\cite{agashe2023llm}, and MASTER~\cite{gan2025master} design various LLM-based multi-agent frameworks where agents collaborate or decompose sub-tasks to achieve higher accuracy and efficiency. Some researches~\cite{wang2025selfevolving} test robustness and generalization of LLM agents by dynamically evolving difficulty of task. While these researches has demonstrated that multi-agent paradigm can efficiently enhance performance on conventional tasks, these benchmarks mainly focus on single-step tasks and centralized orchestration but ignore the organizational structure of multi-agents and multi-turn cooperation tasks. In contrast with conventional benchmarks, Tool-RoCo proposes three multi-robot long-term cooperation tasks.

\newcommand{\cmark}{\ding{51}} 
\newcommand{\xmark}{\ding{55}} 
\begin{table}[h]
\caption{Tool-RoCo and Other Multi-agent Benchmarks}
\label{tab:benchmarks}
\setlength{\tabcolsep}{5pt}
\resizebox{0.47\textwidth}{!}{ 
\begin{tabular}{lllll}
\toprule
\textbf{Benchmark} & \rotatebox{55}{Multi-Turn} & \rotatebox{55}{Tool-Using} & \rotatebox{55}{Reflection} & \rotatebox{55}{Organization}  \\ 
\midrule
ToolBench~\cite{xu2023tool} & \xmark & \cmark & \xmark & \xmark \\
RotBench~\cite{ye2024rotbench} & \xmark & \cmark & \xmark & \xmark \\
\midrule
Collab-Overcook~\cite{sun2025collabovercookedbenchmarkingevaluatinglarge} & \cmark & \xmark  & \cmark & \xmark \\
SMAC~\cite{li2025llm} & \cmark & \xmark  & \xmark & \xmark \\
MultiAgentBench~\cite{wu2023autogenenablingnextgenllm} & \cmark & \xmark & \xmark & \xmark  \\
LLM-Coordination~\cite{agashe2023llm} & \cmark & \xmark & \cmark & \xmark \\
RoCo~\cite{mandi2024roco} & \cmark & \cmark & \xmark & \xmark \\ 
\midrule
Tool-RoCo~\textbf{(Ours)} & \cmark & \cmark & \cmark & \cmark \\ 
\bottomrule
\end{tabular}
}
\end{table}

\par
The second applies LLM to tasks that require coordination to accomplish shared goals, including multi-robot manipulation~\cite{mandi2024roco} and electric game~\cite{li2025llm,sun2025collabovercookedbenchmarkingevaluatinglarge}. Existing studies, such as BudgetMLAgent~\cite{gandhi2024budgetmlagent} and ideation/simulation-focused works~\cite{tanaka2024ideation}, demonstrate that LLM agents can decompose tasks, assign roles, and communicate effectively. 
These benchmarks emphasize role assignment and planning as mechanisms to achieve shared goals. However, most of these systems rely on predefined communication protocols or fixed activation patterns, which limit the evaluation of organization and cooperation in LLMs. Furthermore, the agents in these benchmarks have a fixed action space, and do not have an extensible tools-using set.

\begin{figure*}[t]
  \centering
  \includegraphics[width=1\linewidth]{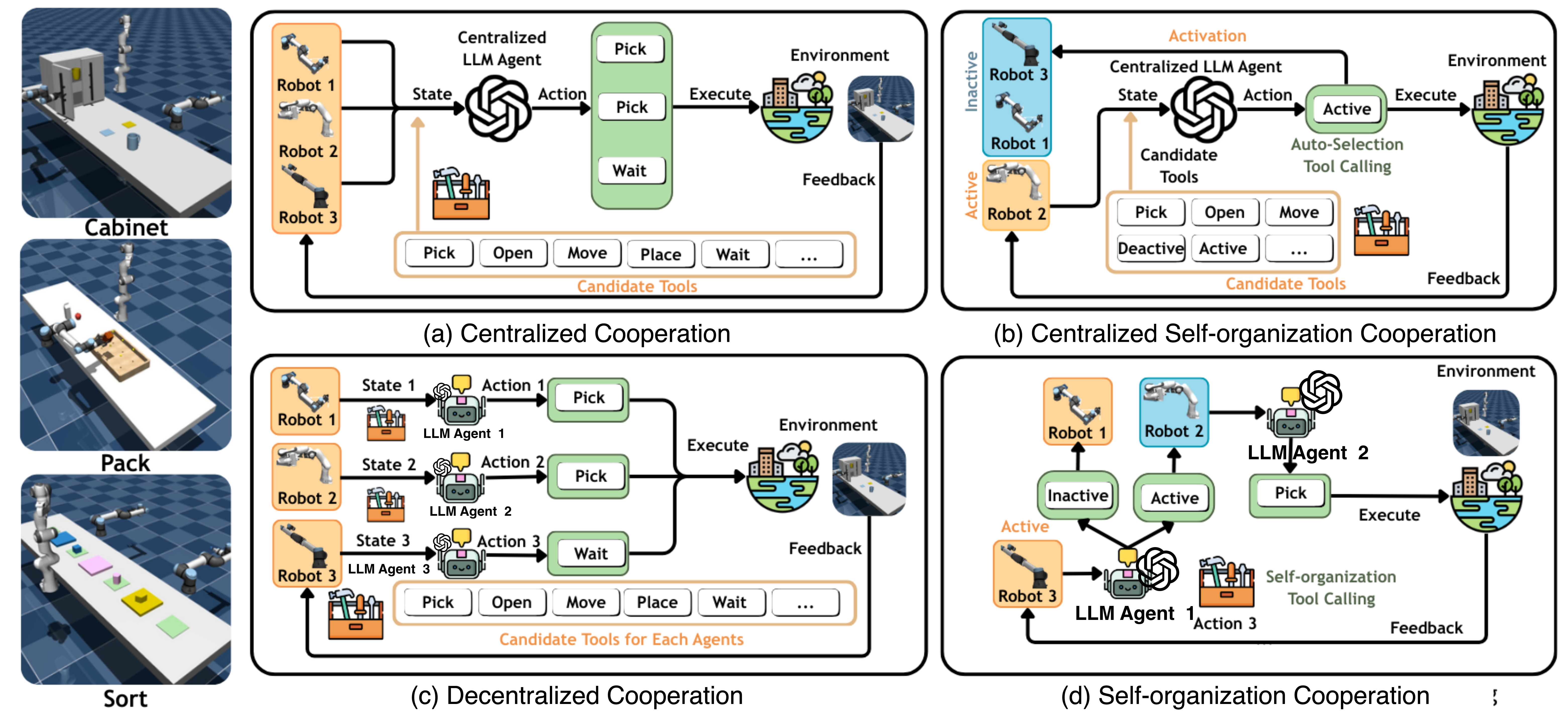}
  \caption{Three tasks and four cooperative paradigms in Tool-RoCo. The paradigms are organized along two dimensions: (i) whether tool selection is centralized by a single LLM agent or distributed across multiple agents, and (ii) whether agents are treated as tools or not. Some other agents that can be activated/deactivated for collaboration. From left(a) to right(d), the paradigms gradually increase in autonomy and complexity: centralized cooperation, centralized self-organization, decentralized cooperation, and self-organized cooperation. }
  \label{fig:method}
\end{figure*}

\subsection{Tool Usage Evaluation and Metrics}
Tool usage~\cite{chen2025toleap}, enabling LLMs to acquire human-like tool-use proficiency for handling complex tasks, has been a critical component in current LLMs. Prior works~\cite{ye2024rotbench,xu2023tool} have measured LLM in tool invocation accuracy, parameter correctness, and analyzed the cost of tool usage~\cite{gandhi2024budgetmlagent}. Although current research mainly focuses on single-agent tool usage or short-term scenarios, tool-using metrics can be effectively used to measure the self-organization and cooperation abilities of LLMs in multi-agent scenarios. Tool-RoCo fills this gap by embedding tool usage in cooperation and treating other agents as tools, thus introducing metrics such as self-organization and cooperation-tool ratio. These metrics enable Tool-RoCo to systematically evaluate organization capability of LLM in long horizontal tasks.

\section{Tool-RoCo}
\subsection{Overview}
\def\AgentSet{\mathcal{A}}
Tool-RoCo is an LLM-based long-term multi-agent benchmark built upon the three tasks in the RoCo environment~\cite{mandi2024roco}: CABINET, PACK, and SORT~(Left side of Fig.\ref{fig:method}). Each task is defined as follows:
\begin{itemize}
    \item \textbf{Cabinet}: ~ Three robots, Alice, Bob, Chad together must take a MUG and a CUP and place them on the correct coasters. The CUP is in the cabinet, and the MUG is on the table.
    \item \textbf{Pack}: ~ Two robots, Alice and Bob, each stand at a different side of the table, and together pack all the grocery items on the table into a bin. 
    \item \textbf{Sort}: ~ Seven panels on the table, ordered left to right: panel1, ..., panel7. They form a straight assembly line, panel1 is close to panel2 and farthest from panel7. There are three cubes, each robot must place their cube on the correct target. There are three robots, each with a limited reach range, this means they can only firstly pick cubes from these panels, and then move or place these cubes on some appropriate middle coordination.
\end{itemize}
\subsection{Dec-
POMDP}
Each task in Tool-RoCo can be formulated as a decentralized partially observable Markov decision process (Dec-POMDP) and a tool-selection process. Dec-POMDP generally
contains a five-tuple ($\AgentSet_t$, $S_t$, $O_t$, $A_t$, $R_t$, $\gamma$,
$P$), where $\AgentSet_t=\{1, \dots, n\}$ is the set of $n$ active agents~(robots) in time $t$,
$S_t$ is the set of local states of each individual agent, $(o^1, \dots,
o^n)\in O_t$ is the set of joint observations, and
$A_t=(a^1,a^2,\dots,a^n)$ represents the joint actions taken by
agents. $R_t =(r^1, \dots, r^n)$ denotes the joint reward of the
environment. $\gamma$ and $P$ denote the decay factor and
probabilistic state transition function, respectively. 
\par

During an episode, LLM-based agents first perceive a semantic local observation $o_t$ of the current environment in a natural language format. Then, agents generate structured responses containing reasoning and selected tools $a_t$ from a tool set. An executor parses and executes these selected tools, then obtains feedback on success or failure, $R_t$, from the environment. This feedback is incorporated into the agent history to adjust and modify the next tool selection at $o_{t+1}$. While this framework allows agents to act in a structured manner, the cooperative and organizational abilities of current LLMs remain limited.
\par
Tool-RoCo provides a standardized API, enabling researchers to easily evaluate different LLMs and multi-agent paradigms without additional engineering effort. Moreover, the agent-as-tool abstraction is task-agnostic, making Tool-RoCo readily extensible to other multi-agent domains such as negotiation, scheduling, and multi-modal robotics. This design aims to establish Tool-RoCo not only as an evaluation benchmark but also as a reusable platform for future multi-robot cooperation tasks.

\subsection{Agent-as-Tool}
Conventional multi-agent benchmarks~\cite{chan2023chatevalbetterllmbasedevaluators, anne2025harnessing} typically provide each agent with a fixed number of allies, and the evaluation quality of cooperation solely depends on the final task result. This outcome obscures the evaluation of agent autonomy and cooperation. In contrast to the current multi-agent benchmark, Tool-RoCo first introduces the \emph{Agent-as-Tool} concept, which enables LLM agents to treat other agents as manipulable tools and add a novel type of tool, called cooperative tools, to the candidate tool set.  
\par
Therefore, the candidate tool set of each agent in Tool-RoCo comprises two categories of tools: common tools and cooperative tools. Common tools are original high-level actions of the agent directly interacting with the environment, such as PICK, OPEN, and PLACE. However, collaborative tools are designed to facilitate interaction and coordination with other agents. Notably, both types of tools are treated at the same hierarchical level of the candidate tool set. This means that in the tool selection process of each round, agents can naturally choose cooperative tools in the same way as common tools, allowing an agent to request assistance from other agents in some complex tasks.
Meanwhile, cooperative tools include deactivation tool like Disconnect~(agent list), which enables an agent to release collaborators if they are no longer needed. 
\par
To facilitate this process, in Tool-RoCo, each agent maintains an active agent pool that records which agents are currently active and which are inactive. When an active tool of cooperative tools is used to connect an agent, the target agent is added to the pool; when a disconnect tool is invoked, the agent is removed. This pool allows each agent to manage dynamic collaboration efficiently and ensures that subsequent tool selections can take into account the current set of active collaborators.
\par
While the Agent-as-Tool and cooperative tools provide a flexible perspective for multi-agent collaboration, achieving decentralized autonomous coordination remains challenging for current LLMs. Decentralized autonomous coordination represents each agent dynamically self-organizing the allies team and disconnecting redundant agents. Therefore, to gradually evaluate autonomy and cooperation capabilities of LLMs, Tool-RoCo defines four progressive cooperative paradigms that gradually increase in complexity: starting from centralized cooperation, through centralized self-organization and decentralized cooperation, and finally arrive at whole self-organization. We explain these four paradigms in the next section.

\subsection{Four Progressive Cooperation Paradigm}

Based on the centralized LLM agent or decentralized multiple LLM agents, and whether other agents are treated as tools, Tool-RoCo defines four progressive paradigms corresponding to increasing levels of agent autonomy. From simplest to hardest, the sequences are Centralized Cooperation, Centralized Self-organization, Decentralized Cooperation, and Self-organization Cooperation. These four paradigms allow us to evaluate the ability of LLM agents systemically.

\begin{table}[h]
\caption{Categories of cooperative paradigm in Tool-RoCo}
\label{tab:categories}
\resizebox{0.47\textwidth}{!}{ 
\begin{tabular}{lll}
\toprule
 Cooperation Paradigm & \multicolumn{1}{c}{\textbf{Centralized LLM}} & \multicolumn{1}{c}{\textbf{Decentralized LLMs}} \\
\midrule
\multicolumn{1}{c|}{\textbf{Agent-not-as-Tool}} & \multicolumn{1}{c|}{Centralized } & \multicolumn{1}{c}{Decentralized }  \\
\midrule
\multicolumn{1}{c|}{\textbf{Agent-as-Tool}} &  \multicolumn{1}{c|}{Centralized Self-organization}  & \multicolumn{1}{c}{Self-organization} \\
\bottomrule
\end{tabular}
}
\end{table}

\subsubsection{Centralized Cooperation}
In the centralized cooperation paradigm, a single LLM-based agent is able to perceive full observability of the environment and allocate tools to all agents as a central policy, effectively orchestrating the entire system. Since an LLM-based agent has access to global information, this paradigm mainly represents the upper bound of task efficiency. Correspondingly, the number of prompt tokens increases in comparison with the decentralized paradigm, as it must encode the states, actions, and all candidate tool sets of all agents in a single prompt. Overall, the centralized cooperation paradigm provides a baseline to measure the basic ability of LLM, including tool selection, parameter filling, and execution validity.
\par
However, this paradigm still makes it difficult to evaluate the cooperative abilities of agents, as all agents are active in every turn and execute together according to a central LLM policy. To address this limitation, the centralized self-organization paradigm is introduced as a medium-level cooperative paradigm.
\subsubsection{Centralized Self-organization}
Next level, the centralized self-organization paradigm in Tool-RoCo retains the central LLM agent but introduces the agent-as-Tool concept into the centralized paradigm, which the central LLM additionally needs to select appropriate agents to activate when selecting tools in each round. Specifically, in the self-organization paradigm, the central LLM agent maintains an active agent pool. This central LLM agent not only assigns tools but also decides which agents should be added to this pool or be deactivated, treating other agents as cooperative tools. 
\par

This requires reflection on redundancy and rational agents allocation, guides the central LLM agent to balance efficiency and attempt to cooperate with other agents. By incorporating activation and deactivation tools, this paradigm provides the first step toward measuring the cooperative ability of the LLM agent. However, it is still constrained by a fully centralized architecture, where collaboration is ultimately decided by a single central LLM agent. This limits the ability to assess whether cooperation can emerge autonomously among distributed agents.

\begin{figure}[t]
    \centering
    \includegraphics[width=1\linewidth]{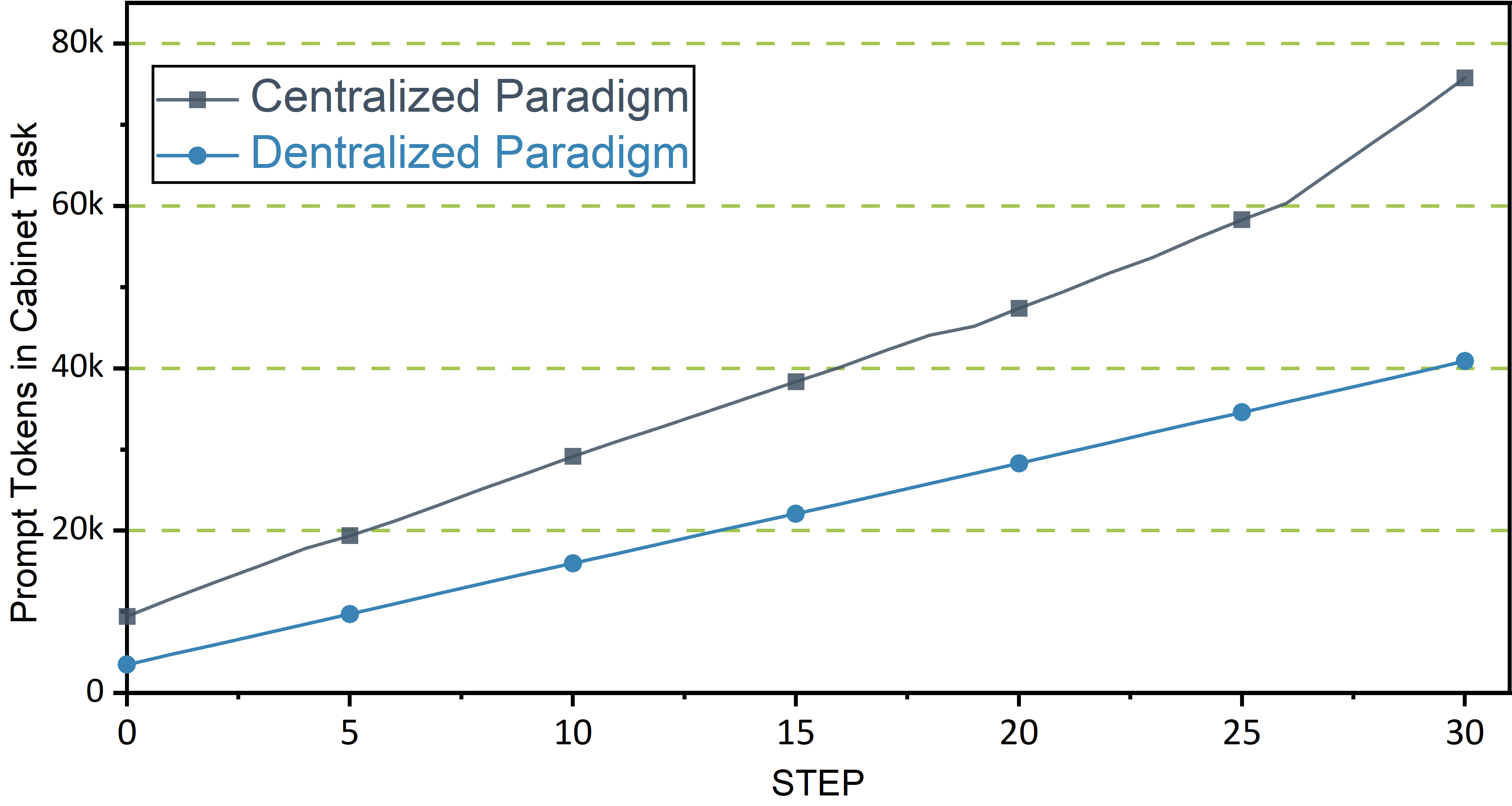}
    \caption{Comparison of prompt token consumption between centralized and decentralized paradigms. LLM agent in Centralized paradigms requires global states, actions, and candidate tool sets of all agents, whereas each agent in decentralized paradigms distributes tool selection, reducing per-agent token consumption. }
    \label{fig:PromptToken}
\end{figure}

\subsubsection{Decentralized Cooperation}

In the decentralized cooperation paradigm, each agent is equipped with its own LLM agent that makes local decisions based on partial observations. Unlike centralized paradigms, there is no global state to guarantee coordination to complete the task. Agents must rely on limited information, which stresses their capacity for local reasoning and cooperative behavior in uncertain situations. This paradigm tests whether LLMs can achieve meaningful coordination without centralized orchestration.
\par
While decentralization evaluates distributed coordination, it still presumes that all agents are initially active and engaged in the task. To extend the boundary, Tool-RoCo introduces the self-organization cooperation paradigm.

\subsubsection{Self-organization Cooperation}
As for Self-organization Cooperation, each agent maintains its own activation pool, and only a single randomly chosen agent is initially active. This agent must autonomously decide when and how to activate additional agents through cooperative tool calls. Once activated, these newly engaged agents can also invoke cooperative tools to activate further or deactivate others, enabling a dynamically self-organized team.
\par
Therefore, this self-organization cooperation paradigm represents the highest level of autonomy. Distributed LLM Agents must not only execute local actions, but also consider when collaboration is needed, which allies to engage, and how to manage other agents activation. By requiring LLM agents to coordinate and adaptively form a team, this paradigm provides a novel criterion for evaluating self-organization and multi-agent collaboration skills in current LLM agents.

\subsection{Self-organization and Cooperative Tool Ratio Metrics}
Although introducing the agent-as-tool concept into the cooperation paradigm, Tool-RoCo explicitly measures the autonomy and cooperative capacity of LLM agents. Therefore, this paper proposes two novel metrics~(Table~\ref{tab:two novel metrics}) that capture different aspects of autonomy and self-organization in cooperation. The first metric is the Cooperative tool ratio of select tools(CT), defined as:
$$ CT = \frac{N_{cooperative}}{N_{tools}}$$
where $N_{cooperative}$ denotes the number of cooperative tool calls~(activating or deactivating other agents) and $N_{tools}$ 
represents the accumulated number of tool calls in all episodes~(Not the size of action space).
This ratio measures how frequently an agent engages others as collaborators or disconnects redundant agents rather than acting in isolation. A higher ratio value indicates stronger recognition of autonomy in cooperation and adaptive leverage of teammates.
\par
The second metric is the Self-Organization ratio~(SO):
$$ SO = \frac{N_{activate}}{N_{cooperative}+\epsilon}$$
Which $N_{active}$ denotes the number of activation calls that explicitly engage other agents in the task. This metric reflects the extent to which agents actively seek assistance from others rather than attempting isolated problem-solving.
\par
The cooperative tool ratio and Self-Organization ratio extend beyond traditional measurements such as tool selection, parameter validation, and execution validity. These metrics offer insights into how LLM agents autonomously initiate, manage, and adapt collaboration, thereby offering a more comprehensive evaluation of emergent cooperation and self-organization in multi-agent systems. In the following section, we adopt these proposed metrics to evaluate the four different cooperation paradigms in the tasks of  Tool-RoCo.

\section{Experiment and Metrics}
\subsection{Experiment Setting}
To evaluate the autonomy and cooperation ability of LLM-based agents, this paper conducts experiments across three representative tasks (\textit{Cabinet}, \textit{Pack}, and \textit{Sort}) in RoCo \cite{mandi2024roco} under four cooperation paradigms separately introduced in Section 3.3. There are multiple agents in each task, and each agent equips a candidate tool set consisting of eight to nine tools to cooperate in achieving the final target. 


This paper runs five independent episodes for each task. Within each episode, agents are allowed to finish the target (take action) in ten turns, with a maximum of five re-planning opportunities. Re-planning opportunities simulate realistic iterative decision-making and enable LLM-based agents to adjust their tool selections based on environment feedback or the actions of other agents. For instance, if a tool selection is invalid or fails, the agent may re-plan its next tool based on the environment feedback. The experiments were conducted with a total cost of approximately \$110, averaging around \$4 per task.
\par
For each decentralized cooperation and self-organization paradigm, all agents are equipped with the same LLM model, reflecting the LLM agent ability to handle varying levels of autonomy and coordination, rather than variations in model capabilities. All four cooperation paradigms are evaluated under this consistent experimental setting, enabling a systematic comparison of LLM autonomy and collaborative behavior.

\definecolor{mylightgray}{gray}{0.9}
\begin{table*}[t]

    \centering
    \scriptsize
    \renewcommand{\arraystretch}{1.2}
    \caption{Performance of various LLMs in three cooperative tasks (\textit{Cabinet}, \textit{Pack}, and \textit{Sort}) of Tool-RoCo.}
    \label{tab:main_combined_final_modified}
    \setlength{\tabcolsep}{3pt}
     \begin{adjustbox}{width=2\columnwidth}
    \begin{NiceTabular}{c|m{0.5cm}|c|ccc|ccc|ccc|ccc|ccc|ccc}
    \CodeBefore
        \rowcolor{mylightgray}{3}
        \rowcolor{mylightgray}{8}
        \rowcolor{mylightgray}{13}
        \rowcolor{mylightgray}{18}
    \Body
        \toprule
        \multicolumn{3}{c|}{} & \multicolumn{6}{c|}{\textbf{Task: Cabinet}} & \multicolumn{6}{c}{\textbf{Task: Pack}} & \multicolumn{6}{c}{\textbf{Task:Sort}} \\
        
        \cmidrule(lr){4-9} \cmidrule(lr){10-15} \cmidrule(lr){16-21}
        \multicolumn{3}{l|}{\textbf{Model Name}} & 
        \rotatebox{90}{Tool calling/\%} & \rotatebox{90}{Param Val/\%} & \rotatebox{90}{Exec Val/\%} & \rotatebox{90}{Reflec Rate/\%} & \rotatebox{90}{Modifi Rate/\%} & \rotatebox{90}{Win(0/1)} &\rotatebox{90}{Tool calling/\%} & \rotatebox{90}{Param Val/\%} & \rotatebox{90}{Exec Val/\%} & \rotatebox{90}{Reflec Rate/\%} & \rotatebox{90}{Modifi Rate/\%} & \rotatebox{90}{Win(0/1)} &
        \rotatebox{90}{Tool calling/\%} & \rotatebox{90}{Param Val/\%} & \rotatebox{90}{Exec Val/\%} & \rotatebox{90}{Reflec Rate/\%} & \rotatebox{90}{Modifi Rate/\%} & \rotatebox{90}{Win(0/1)}
        \\
        \midrule
        \multicolumn{22}{c}{\textit{Centralized Cooperation Paradigm}} \\
        \midrule
        \multicolumn{3}{c|}{\textbf{GPT-4o-mini}} & 19.40 & 10.52 & 9.40 & 16.80 & 7.25 & \cmark & 1.35 & 0 & 0 & 2.17 & 0 & \xmark & 20.57 & 10.19 & 10.19 & 25.67 & 6.04 & \cmark\\
        \multicolumn{3}{c|}{\textbf{GPT-5-mini}} & 10.91 & 9.59 & 9.23 & 14.27 & 6.47  & \xmark & 42.76 & 31.72 & 22.69 & 51.38 & 26.55 & \cmark & 57.89 & 48.03 & 45.39 & 49.34 & 22.37 & \cmark \\
        \multicolumn{3}{c|}{\textbf{GPT-4.1}} & 73.72 & 71.53 & 71.53 & 36.49 & \textbf{28.47}  & \cmark & 81.81 & 48.29 & 29.76 & \textbf{74.39} & \textbf{28.29} & \cmark & 81.82 & 63.64 & 58.18 & \textbf{63.64} & \textbf{34.55} & \cmark\\
        \multicolumn{3}{c|}{\textbf{GPT-5}} & \textbf{93.04} & \textbf{78.31} & \textbf{75.90} & \textbf{43.37} & 19.88 & \cmark & \textbf{82.11} & \textbf{64.22} & \textbf{51.61} & 63.49 & 23.46 & \cmark & \textbf{99.39} & \textbf{88.55} & \textbf{85.54} & 24.69 & 10.84 & \cmark\\
        \midrule
        
        \multicolumn{22}{c}{\textit{Centralized self-organized Cooperation Paradigm}} \\
        \midrule
        \multicolumn{3}{c|}{\textbf{GPT-4o-mini}} & 48.53 & 44.10 & 42.89 & 45.80 & 23.19 & \xmark & 22.32 & 2.20 & 2.20 & 28.63 & 4.30 & \xmark & 54.01 & 11.68 & 11.68 & 19.34 & 5.84 & \xmark \\
        \multicolumn{3}{c|}{\textbf{GPT-5-mini}} & 18.23 & 18.23 & 18.23 & 26.69 & 10.53 & \cmark & 57.33 & 47.07 & 36.48 & 56.51 & 36.48 & \xmark& 58.69 & 54.35 & 54.35 & 21.01 & 10.87 & \xmark\\
        \multicolumn{3}{c|}{\textbf{GPT-4.1}} & 70.03 & 56.10 & 56.10 & \textbf{70.00} & \textbf{20.97} & \cmark & 96.92 & 36.41 & 25.13 & \textbf{84.13} & \textbf{27.69} & \cmark & 92.18 & 70.31 & 62.50 & \textbf{69.53} & \textbf{29.69} & \cmark\\
        \multicolumn{3}{c|}{\textbf{GPT-5}} & \textbf{99.91} & \textbf{86.36} & \textbf{86.36} & 36.36 & 13.64 & \cmark & \textbf{99.32} & \textbf{97.67} & \textbf{72.09} & 38.37 & 2.33 & \cmark & \textbf{100} & \textbf{98.37} & \textbf{98.37} & 20.62 & 8.43 & \cmark\\
        \midrule
        
        \multicolumn{22}{c}{\textit{Decentralized Cooperation Paradigm}} \\
        \midrule
        \multicolumn{3}{c}{\textbf{GPT-4o-mini}} & 3.42 & 1.35 & 1.35 & 4.49 & 0 & \xmark & 9.52 & 5.71 & 5.71 & 8.57 & 0 & \xmark  & 9.52 & 5.71 & 10.19 & 8.57 & 3.35 & \xmark \\
        \multicolumn{3}{c|}{\textbf{GPT-5-mini}} & 10.04 & 8.82 & 8.80 & 43.16 & 8.95 & \xmark  & 19.10 & 13.86 & 13.86 & 28.09 & 11.98 & \xmark  & 26.25 & 16.67 & 16.67 & 38.33 & 10.00 & \xmark \\
        \multicolumn{3}{c|}{\textbf{GPT-4.1}} & 62.75 & 40.20 & \textbf{27.45} & \textbf{78.43} & \textbf{32.35} & \xmark  & 77.78 & 44.45 & 13.33 & \textbf{71.11} & 28.89 & \xmark  & 69.35 & 38.70 & 38.70 & \textbf{67.74} & \textbf{32.25} & \xmark \\
        \multicolumn{3}{c|}{\textbf{GPT-5}} & \textbf{99.27} & \textbf{48.91} & 21.17 & 61.75 & 21.90 & \xmark  & \textbf{83.89} & \textbf{53.02} & \textbf{36.91} & 70.47 & 30.87 & \xmark  & \textbf{94.17} & \textbf{70.00} & \textbf{70.00} & 43.33 & 21.67 & \xmark \\
        \midrule
        
        \multicolumn{22}{c}{\textit{Self-organized Cooperation Paradigm}} \\
        \midrule
        \multicolumn{3}{c}{\textbf{GPT-4o-mini}} & 2.34 & 0 & 0 & 3.07 & 0 & \xmark  & 1.35 & 0 & 0 & 0 & 2.68 & \xmark  & 1.78 & 0 & 0 & 8.93 & 0 & \xmark \\
        \multicolumn{3}{c|}{\textbf{GPT-5-mini}} & 11.45 & 8.21 & 8.21 & 78.45 & 26.15 & \xmark  & 11.37 & 9.03 & 4.68 & 32.11 & 9.69 & \xmark  & 10.25 & 7.06 & 7.06 & 22.61 & 6.72 & \xmark \\
        \multicolumn{3}{c|}{\textbf{GPT-4.1}} & 46.06 & 23.62 & 17.32 & \textbf{78.46} & \textbf{26.15} & \xmark  & 81.99 & 42.24 & 40.37 & \textbf{57.76} & \textbf{17.39} & \xmark  & 50.94 & 32.08 & 32.08 & 49.06 & 25.00 & \xmark \\
        \multicolumn{3}{c|}{\textbf{GPT-5}} & \textbf{80.77} & \textbf{41.54} & \textbf{30.00} & 58.27 & 22.05 & \cmark & \textbf{87.76} & \textbf{84.56} & \textbf{83.67} & 19.39 & 14.29 & \cmark  & \textbf{88.24} & \textbf{60.29} & \textbf{53.68} & \textbf{62.50} & \textbf{26.47} & \xmark \\
        \bottomrule
    \end{NiceTabular}
\end{adjustbox}
\end{table*}

\subsection{Evaluation Metrics}
Although Tool-RoCo builds two novel metrics, Cooperative Tool Ratio (CT) and Self-Organization (SO), to quantify the autonomy and cooperation ability of LLM in multi-agent cooperative tasks, these two metrics are insufficient to precisely reflect the ability of LLM. Therefore, except for CT and SO metrics, Tool-RoCo also  incorporates fundamental metrics for tool calling, enabling a progressive evaluation from basic proficiency to complex cooperative capability:
\begin{itemize}
    \item \textbf{Tool Calling (\%)}: The proportion of successful tool selections when LLM agents take actions, measuring the basic ability of an agent to generate a structured response and use tools correctly. (e.g., PICK versus PCIK(wrong tool name))
    \item \textbf{Parameter Validation (\%)}: The ratio of tool calls with correctly formatted parameters, indicating whether agents provide inputs of the expected type (e.g., string versus float).
    \item \textbf{Execution Validity (\%)}: The proportion of tool calls that successfully execute without violating environment constraints or causing conflicts with other agents, capturing the reachability of execution.
    \item \textbf{Reflection Rate (\%)}: Whether agents can adjust their plans based on environmental feedback, indicating adaptability and reflection in long-term decision-making.
     \[
    \resizebox{0.47\textwidth}{!}{$
    ~~\textbf{Reflection Rate} = \frac{\sum^T_tR(r_{t-1},r_{t})}{N_{tools}} ~~ R = 1 ~~ if ~~r_{t-1 } \neq r_{t} ~~ else ~~ 0$}
    \]
    \item \textbf{Modification Rate (\%)}: Similar with \textbf{Reflection Rate}, but Modification rate just applied into higher received reward $r_t$ compared with previous reward $r_{t-1}$, capturing adaptability of LLM action.
    \[
    \resizebox{0.47\textwidth}{!}{$
    \textbf{Modification Rate} =
    \frac{\sum_{t=1}^{T} M(r_{t-1}, r_t)}{N_{\text{tools}}},\ 
    M = \begin{cases}
    1, & r_t > r_{t-1} \\
    0, & \text{otherwise}
    \end{cases}
    $}
    \]

    \item \textbf{Win (0/1)}: A binary metric indicates whether the agent team successfully completes the task within limited turns.

\end{itemize} 

By combining these basic metrics and proposed novel metrics, Tool-RoCo forms a hierarchical evaluation framework that assesses both individual tool proficiency and cooperative behaviors of LLM agents. This benchmark allows us to analyze the performance of LLM agents progressively across four different cooperation paradigms in the next section.

\section{Result}
Our experiments evaluate four various LLM models~(GPT4o-mini\cite{hurst2024gpt}, GPT4.1\cite{mao2024gpteval}, GPT5-mini, GPT5\cite{zhang2023complete}) across three representative tasks (\textit{Cabinet}, \textit{Pack}, and \textit{Sort}) within four cooperation paradigms in Tool-RoCo framework. To provide a comprehensive evaluation for LLMs in multi-agent cooperative task, results are formalized from two perspectives: (1) fundamental tool-usage and execution metrics that measure baseline proficiency of LLM agents (Table~\ref{tab:main_combined_final_modified}); and (2) two proposed novel cooperative metrics, Cooperative Tool Ratio (CT) and Self-Organization (SO), that capture autonomy and emergent coordination among agents (Table~\ref{tab:two novel metrics}).
\par

\subsection{Fundamental Tool Usage and Reflection Performance}
Table~\ref{tab:main_combined_final_modified} illustrates the fundamental capability of LLMs agents in Tool-RoCo, including tool usage, reflection, modification, and win under four cooperation paradigms. The results reveal a consistent tendency that larger LLMs exhibit stronger structured tool usage compared to smaller ones. For example, in the \textit{Cabinet} task under the centralized paradigm, GPT-5 achieves over 93.04\% tool-calling success with 75.90\% execution validity, whereas GPT-4o-mini remains below 19.40\% and 9.40\%, respectively. This result illustrates a common: larger models enable the invocation of tools effectively and progress toward task goals.
\par
Beyond basic tool metrics, reflection and modification behaviors further differentiate models. Compared with smaller LLMs such as GPT-4o-mini(16.8\%) and GPT-5-mini(14.27\%), larger models like GPT-4.1(36.49\%) and GPT-5(43.37\%) demonstrate stronger reflection and modification capabilities. This suggests that agents within a larger model can revise past failures and modify the next selected tools.
\par
However, notably, although the latest GPT-5 achieves higher tool calling, execution validity, and win, it conducts lower reflection and modification in contrast with GPT-4.1. This suggests that  reflection and modification of GPT-4.1 is better than the latest GPT-5

\par
In addition, the impact of cooperation paradigms is equally evident. Centralized Cooperation Paradigm yields the most stable tool usage, as coordination is managed by a centralized LLM agent. 
Due to the decreasing state space in the Centralized Self-Organized Cooperation Paradigm, all agents are enabled to perform better tool calling and parameter validation. In contrast, decentralized and self-organized paradigms expose the brittleness of smaller models, where parameter validity and execution rates collapse; the current decentralized paradigm is relatively difficult for current LLM agents. Notably, GPT-5 still maintains resilience even under decentralized settings, achieving high tool-calling accuracy in tasks, under distributed reasoning and local tool usage capabilities.  
\par
Overall, these results provide a first-step milestone in LLM performance. While larger models achieve superior performance in structured tool usage, success depends not only on tool calling accuracy but also on the interplay between reflective adaptation and the complexity of the paradigm. This motivates Tool-RoCo to move beyond fundamental tool metrics and examine higher-level cooperative measures, such as cooperative tool ratio (CT) and self-organization (SO), which more directly capture autonomy and coordination in multi-agent systems.  

\subsection{Autonomy and Self-organization among Multiple LLM Agents}

\definecolor{mylightgray}{gray}{0.9}
\begin{table}[t]
\caption{Self-organization (SO) and cooperative tool ratio (CT) of different LLM agents in Tool-RoCo under \textit{Centralized Self-organized Cooperation} and \textit{Self-organized Cooperation} two paradigms. }
\label{tab:two novel metrics}
\setlength{\tabcolsep}{5pt}
\begin{adjustbox}{width=1\columnwidth}
\begin{tabular}{lllllll}
\toprule
\multirow{2}{*}{LLM} & \multicolumn{2}{c|}{\textbf{Cabinet}} & \multicolumn{2}{c|}{\textbf{PACK}} & \multicolumn{2}{c}{\textbf{SORT}} \\
\cmidrule(lr){2-3} \cmidrule(lr){4-5} \cmidrule(lr){6-7}
  & CT(\%) & SO(\%) & CT(\%) & SO(\%)  & CT(\%) & SO(\%) \\ 
\midrule
 \rowcolor{mylightgray}  \multicolumn{7}{c}{Centralized self-organized Cooperation Paradigm} \\
 \midrule
\textbf{GPT-4o-mini} & 1.73 & 90.91 & 0 & 0 & 0.45 & 100\\
\textbf{GPT-5-mini} & 2.38 & 92.59 & 1.78 & 100 & 1.69 & 100\\
\textbf{GPT-4.1} &8.19& 73.91 & 5.69 & 84.32 & 4.47 & 100\\
\textbf{GPT-5} &9.28 & 72.73 & 2.08 & 50.00 & 2.06 & 84.61\\
\midrule
 \rowcolor{mylightgray}  \multicolumn{7}{c}{Self-organized Cooperation Paradigm} \\
 \midrule
 \textbf{GPT-4o-mini} & 0 & 0 & 0 & 0 & 1.78 & 100\\
\textbf{GPT-5-mini} &  4.17 & 100 & 0 & 0 & 1.72  & 100\\
\textbf{GPT-4.1} & 17.16 & 100 & 1.19 & 100 & 2.78 & 100\\
\textbf{GPT-5} & 26.31 & 100 & 5.73 & 100 & 10.81& 100\\
\bottomrule
\end{tabular}
\end{adjustbox}
\end{table}

According to the agent-as-tool concept, Table \ref{tab:two novel metrics} evaluates the autonomy and self-organization of current LLM agents under just two cooperation paradigms: \textbf{Centralized Self-Organized Cooperation Paradigm} and \textbf{Self-Organized Cooperation Paradigm}. Because only these two paradigms treat other agents as callable tools, maintaining an active agent pool for cooperative behavior, and thereby can evaluate autonomy and self-organization of current LLM-based agents.
\par
In \textbf{Centralized Self-Organized Cooperation paradigm}, smaller models such as GPT-4o-mini and GPT-5-mini achieve very low cooperative tool ratios (CT), ranging from 0\% to 2.38\%. This indicates that while smaller models rarely attempt to collaborate with other agents or even never request assistance at all~(GPT-4o-mini in PACK). In contrast, larger size LLM agents~(GPT-4.1 and GPT-5) demonstrate a modest improvement in CT, ranging from 2.08\% to 9.28\%, reflecting that larger LLM agents consider selecting cooperation-relevant tools and have a limited cooperative ability.
\par
Interestingly, smaller size LLM-based agents exhibit high self-organization (SO) scores in some tasks, which suggests that agents prefer activating other agents. By contrast, GPT-4.1 and GPT-5 achieve lower SO, and these results do not demonstrate that there exists a higher level of autonomy in small LLM because this suggests that smaller-sized LLM-based agents cannot discern when to invoke assistance and when to proceed independently.

\par
Under \textbf{Self-Organized Cooperation paradigm}, Table \ref{tab:two novel metrics} reveals a significant distance between small and large models. GPT-4o-mini and GPT-5-mini exhibit low CT~(most values are zeros) across most tasks, indicating that autonomous coordination within the local state in a decentralized paradigm is highly challenging for smaller LLM-based agents. In contrast, GPT-4.1 and GPT-5 achieve significantly higher CT, demonstrating that larger LLMs already have a limited collaboration with other agents. At the same time, SO is extremely high under a decentralized state, suggesting that current LLM agents consistently activate others but lack the ability to determine when to disengage them, revealing a fundamental limitation in current LLM agents within cooperative tasks.
\par
Overall, these experiments in autonomy and self-organization illustrate that the ability of current LLM-based agents to self-organize is insufficient for multi-agent cooperation. High self-organized (SO) values do not always translate into meaningful contribution (CT) toward task completion. Smaller LLM-based agents are reluctant to seek assistance, while larger models can partially increase this gap but still struggle with constrained self-origination ability. Regardless of scale, LLMs agent always has a high SO value, suggesting that they are unable to balance activating versus deactivating agents for cooperation. These results highlight the need for prompting self-organization ability and coordination, motivating future work in improving multi-agent tool-use proficiency.

\section{Conclusion}
This study introduces Tool-RoCo, a long-term multi-robot framework for evaluating autonomy and cooperative behavior in LLM-based multi-agent systems. Tool-RoCo comprises three representative tasks involving different numbers of agents and four cooperation paradigms that cover progressively increasing levels of cooperative complexity. Meanwhile, Tool-RoCo first introduced agent-as-tool into LLM-based multi-agent systems and proposed two novel metrics: Cooperative Tool Ratio (CT) and Self-Organization (SO), combining with fundamental tool usage metrics to provide a comprehensive and hierarchical evaluation for current LLM-based agents. 
\par
Our experiments across three representative tasks and four cooperation paradigms reveal that an advanced LLM-based agent, such as GPT-5, exhibits a strong tool proficiency and high autonomous collaboration.  While LLM-based agents demonstrate an initial cooperative ability by requesting assistance from others, they generally fail to deactivate agents once these agents are no longer required. This limitation causes all activated agents to remain persistently active, resulting in substantial token overhead. These findings highlight both the promise and the limitations of current LLM-based multi-agent systems, establishing Tool-RoCo as a systematic benchmark for future research on autonomy and cooperation in LLM agents.
\section{Further work}
Although Tool-RoCo introduces CT and SO as novel metrics for quantifying cooperation and self-organization, these two measures are insufficient to capture all the characteristics of multi-agent collaboration. Meanwhile, current cooperation tool usage just contains two naive tools and cannot contains all organizational structures of multi-agents. Future research should extend the cooperation tool set and the evaluation with fine-grained criteria, such as temporal efficiency of cooperation and redundancy in agent activation. 
\par
Moreover, Tool-RoCo currently acts as the only benchmark. Our future research will integrate Tool-RoCo into a training environment, enabling reinforcement learning fine-tuning or curriculum learning to enhance cooperative and organizational abilities of LLM agents. With the training method and advanced multi-agent cooperative benchmark, Tool-RoCo will be critical for developing general LLM agents that have robust organization and collaboration capabilities with others.

\bibliography{aaai2026}

@misc{xu2023tool,
      title={On the Tool Manipulation Capability of Open-source Large Language Models}, 
      author={Qiantong Xu and Fenglu Hong and Bo Li and Changran Hu and Zhengyu Chen and Jian Zhang},
      year={2023},
      eprint={2305.16504},
      archivePrefix={arXiv},
      primaryClass={cs.CL}
}

@article{zhang2023complete,
  title={A complete survey on generative ai (aigc): Is chatgpt from gpt-4 to gpt-5 all you need?},
  author={Zhang, Chaoning and Zhang, Chenshuang and Zheng, Sheng and Qiao, Yu and Li, Chenghao and Zhang, Mengchun and Dam, Sumit Kumar and Thwal, Chu Myaet and Tun, Ye Lin and Huy, Le Luang and others},
  journal={arXiv preprint arXiv:2303.11717},
  year={2023}
}

@misc{sun2025collabovercookedbenchmarkingevaluatinglarge,
      title={Collab-Overcooked: Benchmarking and Evaluating Large Language Models as Collaborative Agents}, 
      author={Haochen Sun and Shuwen Zhang and Lujie Niu and Lei Ren and Hao Xu and Hao Fu and Fangkun Zhao and Caixia Yuan and Xiaojie Wang},
      year={2025},
      eprint={2502.20073},
      archivePrefix={arXiv},
      primaryClass={cs.CL},
      url={https://arxiv.org/abs/2502.20073}, 
}

@inproceedings{gandhi2024budgetmlagent,
  title={Budgetmlagent: A cost-effective llm multi-agent system for automating machine learning tasks},
  author={Gandhi, Shubham and Patwardhan, Manasi and Vig, Lovekesh and Shroff, Gautam},
  booktitle={Proceedings of the 4th International Conference on AI-ML Systems},
  pages={1--9},
  year={2024}
}

@article{agashe2023llm,
  title={Llm-coordination: evaluating and analyzing multi-agent coordination abilities in large language models},
  author={Agashe, Saaket and Fan, Yue and Reyna, Anthony and Wang, Xin Eric},
  journal={arXiv preprint arXiv:2310.03903},
  year={2023}
}

@inproceedings{mandi2024roco,
  title={Roco: Dialectic multi-robot collaboration with large language models},
  author={Mandi, Zhao and Jain, Shreeya and Song, Shuran},
  booktitle={2024 IEEE International Conference on Robotics and Automation (ICRA)},
  pages={286--299},
  year={2024},
  organization={IEEE}
}

@misc{wang2025capabilitiesgpt5multimodalmedical,
      title={Capabilities of GPT-5 on Multimodal Medical Reasoning}, 
      author={Shansong Wang and Mingzhe Hu and Qiang Li and Mojtaba Safari and Xiaofeng Yang},
      year={2025},
      eprint={2508.08224},
      archivePrefix={arXiv},
      primaryClass={cs.CL},
      url={https://arxiv.org/abs/2508.08224}, 
}

@inproceedings{mao2024gpteval,
  title={GPTEval: A Survey on Assessments of ChatGPT and GPT-4},
  author={Mao, Rui and Chen, Guanyi and Zhang, Xulang and Guerin, Frank and Cambria, Erik},
  booktitle={Proceedings of the 2024 Joint International Conference on Computational Linguistics, Language Resources and Evaluation (LREC-COLING 2024)},
  pages={7844--7866},
  year={2024}
}

@article{hurst2024gpt,
  title={Gpt-4o system card},
  author={Hurst, Aaron and Lerer, Adam and Goucher, Adam P and Perelman, Adam and Ramesh, Aditya and Clark, Aidan and Ostrow, AJ and Welihinda, Akila and Hayes, Alan and Radford, Alec and others},
  journal={arXiv preprint arXiv:2410.21276},
  year={2024}
}

@article{anne2025harnessing,
  title={Harnessing language for coordination: A framework and benchmark for llm-driven multi-agent control},
  author={Anne, Timoth{\'e}e and Syrkis, Noah and Elhosni, Meriem and Turati, Florian and Legendre, Franck and Jaquier, Alain and Risi, Sebastian},
  journal={IEEE Transactions on Games},
  year={2025},
  publisher={IEEE}
}

@misc{chan2023chatevalbetterllmbasedevaluators,
      title={ChatEval: Towards Better LLM-based Evaluators through Multi-Agent Debate}, 
      author={Chi-Min Chan and Weize Chen and Yusheng Su and Jianxuan Yu and Wei Xue and Shanghang Zhang and Jie Fu and Zhiyuan Liu},
      year={2023},
      eprint={2308.07201},
      archivePrefix={arXiv},
      primaryClass={cs.CL},
      url={https://arxiv.org/abs/2308.07201}, 
}

@article{
chen2025solving,
title={Solving Multi-agent Path Finding as an {LLM} Benchmark: How, How Good and Why},
author={Weizhe Chen and Sven Koenig and Bistra Dilkina},
journal={Transactions on Machine Learning Research},
issn={2835-8856},
year={2025},
url={https://openreview.net/forum?id=8hAxEFRVQT},
note={}
}

@inproceedings{wang2025selfevolving,
  title={Benchmark Self‑Evolving: A Multi-Agent Framework for Dynamic LLM Evaluation},
  author={Wang, Y. and Long, H. and Fan, J. and Huang, Z. and Wei, X.},
  booktitle={COLING},
  year={2025},
  url={https://aclanthology.org/2025.coling-main.223/}
}

@inproceedings{gan2025master,
  title={MASTER: A Multi-Agent System with LLM Specialized MCTS},
  author={Gan, X. and others},
  booktitle={NAACL},
  year={2025},
  url={https://aclanthology.org/2025.naacl-long.476/}
}

@inproceedings{tanaka2024ideation,
  title={Research on Ideation Applications Using LLM-based Multi-agent Systems and Idea Evaluation Methods},
  author={Tanaka, Y. and others},
  booktitle={JSAI},
  year={2024},
  url={https://www.jstage.jst.go.jp/article/pjsai/JSAI2024/0/JSAI2024_4G3GS205/_article/-char/en}
}

@article{lei2024macm,
  title={Macm: Utilizing a multi-agent system for condition mining in solving complex mathematical problems},
  author={Lei, Bin and Zhang, Yi and Zuo, Shan and Payani, Ali and Ding, Caiwen},
  journal={Advances in Neural Information Processing Systems},
  volume={37},
  pages={53418--53437},
  year={2024}
}

@inproceedings{dong2024multi,
  title={The multi-agent system based on llm for online discussions},
  author={Dong, Yihan},
  booktitle={Proceedings of the 23rd International Conference on Autonomous Agents and Multiagent Systems},
  pages={2731--2733},
  year={2024}
}

@inproceedings{zhao2025mage,
  title={Mage: A multi-agent engine for automated rtl code generation},
  author={Zhao, Yujie and Zhang, Hejia and Huang, Hanxian and Yu, Zhongming and Zhao, Jishen},
  booktitle={2025 62nd ACM/IEEE Design Automation Conference (DAC)},
  pages={1--7},
  year={2025},
  organization={IEEE}
}

@inproceedings{
chen2024agentverse,
title={AgentVerse: Facilitating Multi-Agent Collaboration and Exploring Emergent Behaviors},
author={Weize Chen and Yusheng Su and Jingwei Zuo and Cheng Yang and Chenfei Yuan and Chi-Min Chan and Heyang Yu and Yaxi Lu and Yi-Hsin Hung and Chen Qian and Yujia Qin and Xin Cong and Ruobing Xie and Zhiyuan Liu and Maosong Sun and Jie Zhou},
booktitle={The Twelfth International Conference on Learning Representations},
year={2024},
url={https://openreview.net/forum?id=EHg5GDnyq1}
}

@article{li2025llm,
  title={LLM-guided decision-making toolkit for multi-agent reinforcement learning},
  author={Li, Zhemin and Zhang, Ruobing and Wang, Zhengming and Xie, Zheng and Song, Yiping},
  journal={Neurocomputing},
  volume={638},
  pages={130105},
  year={2025},
  publisher={Elsevier}
}

@inproceedings{ye2024rotbench,
  title={RoTBench: A Multi-Level Benchmark for Evaluating the Robustness of Large Language Models in Tool Learning},
  author={Ye, Junjie and Wu, Yilong and Gao, Songyang and Huang, Caishuang and Li, Sixian and Li, Guanyu and Fan, Xiaoran and Zhang, Qi and Gui, Tao and Huang, Xuanjing},
  booktitle={EMNLP},
  year={2024}
}

@article{chen2025toleap,
  title={ToLeaP: Rethinking Development of Tool Learning with Large Language Models},
  author={Chen, Haotian and Song, Zijun and Niu, Boye and Zhang, Ke and Ou, Litu and Lu, Yaxi and Zhang, Zhong and Cong, Xin and Lin, Yankai and Liu, Zhiyuan and others},
  journal={arXiv preprint arXiv:2505.11833},
  year={2025}
}

@article{zhu_multiagentbench_2025,
  title={Multiagentbench: Evaluating the collaboration and competition of llm agents},
  author={Zhu, Kunlun and Du, Hongyi and Hong, Zhaochen and Yang, Xiaocheng and Guo, Shuyi and Wang, Zhe and Wang, Zhenhailong and Qian, Cheng and Tang, Xiangru and Ji, Heng and others},
  journal={arXiv preprint arXiv:2503.01935},
  year={2025}
}

@misc{wu2023autogenenablingnextgenllm,
      title={AutoGen: Enabling Next-Gen LLM Applications via Multi-Agent Conversation}, 
      author={Qingyun Wu and Gagan Bansal and Jieyu Zhang and Yiran Wu and Beibin Li and Erkang Zhu and Li Jiang and Xiaoyun Zhang and Shaokun Zhang and Jiale Liu and Ahmed Hassan Awadallah and Ryen W White and Doug Burger and Chi Wang},
      year={2023},
      eprint={2308.08155},
      archivePrefix={arXiv},
      primaryClass={cs.AI},
      url={https://arxiv.org/abs/2308.08155}, 
}

\end{document}